\documentclass[journal=jacsat,manuscript=article]{achemso}

\usepackage[version=3]{mhchem} 
\usepackage[T1]{fontenc}       
\usepackage{graphics}
\usepackage[usenames, dvipsnames]{color}
\usepackage{longtable}
\usepackage{multirow}
\usepackage{makecell}
\usepackage{natmove}
\setkeys{acs}{usetitle = true}
\usepackage{amsmath,achemso}
\usepackage{bm}
\setkeys{acs}{maxauthors = 9}

\author{Duy Khanh Nguyen}
\email{nguyenkhanhphysics2015@gmail.com}
\affiliation {Department of Physics, National Cheng Kung University, Tainan, Taiwan}

\author{Ngoc Thanh Thuy Tran}
\affiliation {Department of Physics, National Cheng Kung University, Tainan, Taiwan}

\author{Thanh Tien Nguyen}
\affiliation {Department of Physics, College of Natural Sciences, Can Tho University, Can Tho City, Vietnam}

\author{Yu-Huang Chiu}
\affiliation{Department of Applied Physics, National Pingtung University, Pingtung, Taiwan}

\author{Ming-Fa Lin}
\email{mflin@mail.ncku.edu.tw}
\affiliation {Department of Physics, National Cheng Kung University, Tainan, Taiwan}

\title[An \textsf{achemso} demo]
  {Essential Properties of Fluorinated Graphene
  and Graphene Nanoribbons}

\abbreviations{DOS, XPS, STM, ARPES, STS}

\begin{document}

\begin{abstract}
 A systematic study is conducted on the fluorination-enriched essential properties of 2D graphene and 1D graphene nanoribbons using the first-principles method. The combined effects, which arise from the significant chemical bonds in C-C, F-C and F-F bonds, the finite-size quantum confinement, and the edge structure, can greatly diversify geometric structures, electronic properties and magnetic configurations. By the detailed analyses, the critical orbital hybridizations in determining the essential properties are accurately identified from the atom-dominated energy bands, the spatial charge distributions, and the orbital-projected density of states. The top-site F-C bonds, with the multi-orbital hybridizations, create the non-uniform buckled honeycomb lattice. There exist the C-, F- and (C, F)-dominated energy bands. Fluorinated graphene belongs either to the p-type metals (with/without the ferromagnetic spin arrangement) or to the large-gap semiconductors (without magnetism), depending on the concentration and distribution of adatoms. Specially, fluorinated graphene nanoribbons, with armchair/zigzag edge, presents five kinds of spin-dependent properties, covering the non-magnetic and ferromagnetic metals, non-magnetic semiconductors, and anti-ferromagnetic semiconductors with/without the spin splitting. The various band-edge states in 2D and 1D systems appear as the rich and unique structures in density of states. Part of theoretical predictions are consistent with the experimental measurements, and the others are worthy of the further examinations. Also, the fluorination-created diverse properties clearly indicate the high potentials in various applications that will be discussed in detail, e.g., electronic and spintronic nanodevices.
\end{abstract}

\section{I. Introduction}

Since graphene was successful produced, there has been numerous studies both theoretically \cite{1} and experimentally \cite{2} to modulate its properties, especially the electronic structures. Adatom doping on the graphene sheet with different types of elements such as oxygen, hydrogen, alumina, or alkali is one of the efficent methods for diversifying the band structure of graphene \cite{3}. The fluorinated graphenes are commonly synthesized using a two-step process, fluorination to obtain fluorinated graphite\cite{4} and then exfoliation to achieve monolayer one.\cite{5} Monolayer F-doped graphene could also be directly synthesized from the chemical reaction method by heating the mixture of graphene sheet and XeF$_2$.\cite{6} 

Graphene nanoribbons (GNRs) \cite{7} have emerged as the ideal nanomaterials to remove the unexpected gapless feature of 2D graphene due to the complex relations of 1D honeycomb lattices, one-atom thicknesses, finite-size quantum confinements and edge structures. GNRs are successfully fabricated by various experimental methods including both top-down and bottom-up schemes.\cite{8} From the geometric side, one of simplest and intuitive methods to create GNRs is cutting of 2D graphene sheet that covers lithographic patterning \cite{9}, graphene etching \cite{10}, sonochemical breaking \cite{11}, metal-catalyzed treatment.\cite{12}. Also, the massive amount of GNRs can be produced by the chemical vapor deposition or unzipping of the multi-walled carbon nanotubes.\cite{13} The essential properties of GNRs materials can lead to many potential applications in nano-electronic \cite{14} and spintronic \cite{15} devices, gas sensor. \cite{16} Furthermore, the special electronic properties of GNRs can be diversified by many different paths such as mechanical strain \cite{17,18}, layer number \cite{19,20}, curved surface \cite{21}, edge passivation \cite{22,23}, stacking configuration \cite{24}, electric \cite{25,26}; magnetic \cite{27,28} fields and adatoms adsorption \cite{29,30}. Chemical adsorption on the planar surface of GNRs is one of the effective routes that can result in the feature-rich properties. There are a lot of previous studies in terms of chemical doped GNRs. For example, the transition metal adatom-doped GNRs induce the spin-splitting metallic 1D energy bands with free electron density \cite{31}. Alkali-adsorbed GNRs in armchair and zigzag edges, respectively, exhibit the nonmagnetic and ferromagnetic metal with high electron density\cite{32}. The 1D semiconducting or metallic band structures are revealed in the ligand-protected aluminum clusters-adsorbed AGNRs.\cite{33} 

Interestingly, fluorine possesses a very strong electron affinity among other halogen or atoms; they are thus expected to present the significant chemical bondings with carbon atoms and greatly diversify the essential properties. The feature-rich electronic properties of 2D graphene and 1D GNRs are investigated in detail for various F adatom-concentrations and distributions by the first-principles calculations. The geometric structures strongly depend on the top site F-C bonds and so do the electronic properties. The atom-dominated energy bands, the orbital-projected DOS, and the spatial charge distributions can be used to comprehend the critical multi-orbital hybridizations in C-C, F-C and F-F bonds which are responsible for the enriched electronic and magnetic properties. The up-to-date theoretical and experimental studies as well as the potential applications are also discussed.

\section{II. Computational methods}

Modeling simulation on computer schemes, providing an important prediction for further experimental studies, has been emphasized as a pioneering method in physics, chemistry and materials sciences in the recent years. A specific system can be fully described by the  quantum mechanic Schr{\"o}dinger equation. However, such equation is impossible to solve for many-body particle systems. Hence, possible approximations are utilized to deal with the problem. Up to now, the first-principles calculations have been considered as an ideal method for materials simulation since the complicated electron-electron and electron-ion interactions are thoroughly characterized by a series of approximations and simplifications. Furthermore, Vienna ab initio simulation package (VASP) \cite{34} computes an approximate solution of many-body Schr{\"o}dinger equation, either within the density functional theory (DFT) to solve the Kohn-Sham equations, or within the Hartree-Fock approximation to solve the Roothaan equations.
In this study, the adatom-enriched essential properties are investigated by the first-principles DFT and VASP calculations. The exchange-correlation energy due to the electron-electron interactions is calculated from the Perdew-Burke-Ernzerhof functional under the gerneralized gradient approximation. \cite{35} The projector-augmented wave pseudopotentials are used to evaluate the electron-ion interations.\cite{36} The wave functions are built from the plane wave with a maximum energy cutoff of 400 eV, in which they are tested to achieve a very high accuracy. All the atoms in the relaxation process are subjected to the adjustments of their positions in order to have an optimized structures with the lowest total ground state energy. And then, the most stable configuration is used as the input parameters for further calculations on the fundamental physical properties.  The interactions between two neighboring images are eliminated by applying the vacuum distance along the \(z=15\) \AA \ and (\(z=15\) \AA \ , \(y=15\) \AA \ ) axes for 2D graphene and 1D graphene nanoribbons, respectively. To fully understand the surface adsorption effects on the magnetism, the spin configurations are taken into account for the fluorine adatoms-adsorbed graphene and graphene nanoribbons.  The first Brillouin zone is sampled in a Gamma scheme along the 2D and 1D periodic directions by \(12 \times 12 \times 1\)
 and \(12 \times 1 \times 1\) k-points for structure relaxation, and then \(100 \times 100 \times 1\) and \(100 \times 1 \times 1\) for accurate evaluations on electronic properties, respectively. Besides, the convergence criterion for energy is set at \(10^{-5}\) eV between two simulation steps, and the maximum Hellmann-Feynman forces acting on each atom is smaller than 0.01 eV/\AA \ during the ionic relaxations.

\section{III. Results and discussions}
\subsection{1. Fluorinated 2D graphene}

The stable adsorption position of F atoms on graphene is determined by comparing the binding energies, in which the lower value 
corresponds to higher stability. The binding energy ($E_b$) of n adsorbed F adatoms is expressed as $E_b$ = ($E_{tot}$ - $E_{gra}$ - n$E_{Fluo}$)/n, where $E_{tot}$, $E_{gra}$, and $E_{Fluo}$ are the total ground state energies of the fluorinated graphene, pristine graphene, and isolated F adatom, respectively. Among the bridge-, hollow- and top-sites, our calculations show that the top-site is the most stable one, in agreement with previous studies. \cite{37} After self-consistent calculations, we found that all the fluorinated graphene systems, with various concentrations and distributions, have stable buckled structures as demonstrated in Fig. 1. The adatom distributions are divided into two main kinds, in one of which F atoms are absorbed at the same sublattice (Figs. 1(a); 1(c)-1(e)), in another of which adsorptions take place at two different sublattices simultaneously (Figs. 1(b) \& 1(f)). The latter is predicted to be more stable than the former because of the lower binding energy, e.g, the 25\% cases (Table 1). The nearest C-C bond lengths in graphene are expanded in the ranges of 1.45$-$1.58 $\mbox\AA$ with the increase of F-concentrations. As to the F-C bond lengths, they are shorten in the ranges of 1.57$-$1.38 $\mbox\AA$ with the increase of F-concentration, indicating the stronger F-C bonds at the high F-concentrations. 

The two-dimensional band structures along the high symmetry points are useful in  examining the electronic properties. Monolayer graphene has a Dirac-cone structure at the K point (Fig. 2(a)), owing to the extended $\pi$ bondings of 2p$_z$ orbitals in a hexagonal sheet. The linear energy bands turn into the parabolic dispersions in the increase of state energy, e.g., the parabolic bands near the  $M$ point. Furthermore, the $\sigma$ bands, which arise from the sp$^2$ bondings, are initiated from the $\Gamma$ point at the deeper energy. When F atoms are adsorbed on graphene, the energy dispersion relations exhibit dramatic changes with the variations in F-concentrations and F-distributions. The linear isotropic Dirac cone of graphene is destroyed at high F-concentration, mainly owing to the strong F-C bonds (discussed later). Under the full fluorination (Fig. 2(b)), a direct energy gap of 3.12 eV appears at the $\Gamma$ point, in agreement with the previous theoretical calculations\cite{38} and experimental measurements.\cite{39} With the decrease of fluorination, the semiconducting or metallic behaviors are mainly determined by the C-dominated energy bands (Figs. 2(c)-2(f)), since the $\pi$-electronic structure (the 2p$_z$-orbital bonding) is gradually recovered. The F-F bonding quickly declines, while the strength of F-C bond is almost the same, as suggested by the weakly dispersive (F,C)-co-dominated bands. The fluorinated systems, as shown in Figs. 2(d) \& 2(e), have the spin-split energy bands across the Fermi level simultaneously; that is, they present the metallic ferromagnetism. Such bands are dominated by carbon atoms (2p$_z$ orbitals), but not F adatoms. The occupied/unoccupied carrier densities in the spin-up and spin-down bands differ from each other (the black and red curves in Figs. 2(d) \& 2(e)). There exist ferromagnetic spin configurations under the same sublattice adsorption, as observed in Figs. 3(a) and 3(b) for 16.7\% and 12.5\% F-concentration, respectively.

The charge density (Figs. 4(a)-4(c)) and the charge density difference (Figs. 4(d) \& 4(e)) can provide very useful information about the chemical bondings. The latter is obtained by subtracting
the charge density of pristine graphene and F adatoms from that of fluorinated graphene system. As F atoms are adsorbed on graphene, their orbitals have strong hybridizations with those of passivated C, as seen from the red region enclosed by the dashed black rectangles (Figs. 4(b) and 4(c)). Compared with pristine graphene (Fig. 4(a)), such strong F-C bonds lead to the deformed $\pi$ bonds and thus the serious distortion of the Dirac-cone structures (Fig. 2(a)). Between two non-passivated C atoms, $\rho$ shows a strong $\sigma$ bonding (black rectangles), which is slightly reduced after the formation of F-C bonds. There also exist the significant F-F bonds (purple rectangles) where the adatoms are sufficiently close to each other. These can create the (F,C)-co-dominated energy bands, the F-induced energy bands, the more complicated $\pi$ and $\sigma$ bands, and the seriously distorted Dirac cone (Figs. 2(b)-2(f)). The obvious spatial distribution variations on ${(y, z)}$ and ${(x, z)}$ planes clearly illustrate the multi-orbital hybridizations of (2p$_x$, 2p$_y$, 2p$_z$) in F-C bonds. The sp$^3$ bonding resulted in a buckled graphene structure is indicated from the deformations of the $\sigma$ bonding in the nearest C atom (orange arrows) and the $\pi$ bonding in the next-nearest one (purple arrows). Between the non-passivated C atoms, the $\pi$  and $\sigma$ bondings almost remain the same as pristine graphene. Only slightly deformations are presented as in the dashed purple rectangles (Fig. 4(d)).

The orbital-projected DOS can clearly illustrate the orbital hybridizations in F-F, C-F and C-C bonds. For pristine graphene, the low-lying structures, as shown in the upper inset of Fig. 5(a), are caused by the 2p$_z$-2p$_z$ bondings among C atoms (red curve). DOS presents a linear energy dependence near ${E=0}$ and vanishes there, illustrating the characteristic of a zero-gap semiconductor. The $\pi$ and $\pi^*$ logarithmic-form peaks arise from the saddle points of parabolic bands near the M point (the dashed curve in Fig. 2(a)). Moreover, the shoulder structures, which are associated with the (2p$_x$, 2p$_y$) orbitals, come to exist at deeper energy (${\sim\,-3}$ eV). They correspond to the extreme band-edge states of parabolic dispersions near the $\Gamma$ point (Fig. 2(a)). The lower-energy DOS is dramatically altered after fluorination. Under the full fluorination (Fig. 5(b)), the V-shaped structure and two prominent symmetric peaks are absent, since the overall F-C interactions thoroughly destroy the $\pi$ bondings of graphene. Instead, there exist an energy gap  of 3.12 eV centered at $E_F$ and several F-dominated special structures, a result of the F-F bondings. These clearly illustrate the coexistence of strong F-F, C-C and F-C bonds. In addition, the 2p$_z$ orbitals only make contributions at very deep energies (Fig. 5(a)). With the reduce of fluorination, there are more special structures in DOS, as indicated in Figs. 5(b)-5(f), being attributed to the narrower energy widths of F-dependent valence bands, the gradual recovery of carbon $\pi$ bondings (red curves), and the partial contributions of F-2p$_z$ orbitals (blue curves).

The possession of remarkable properties enables fluorinated graphenes in many promising applications. They have been used in lithium-related batteries. \cite{40} Edge-fluorinated graphene nanoplatelets could be used as high performance electrodes for lithium ion batteries and dye-sensitized solar cells. \cite{41} Fluorinated graphenes could also be utilized as an electrode material for supercapacitors. \cite{42} Moreover, the hybridized systems have potential performances in many other fields such as the ink-jet printed technologies, \cite{43} the biological scaffold for promoting neuro-induction of stem cells, \cite{44} and amonia detection. \cite{45}

\subsection{2. Fluorinated 1D graphene nanoribbons}

In this part, the detailed investigation on geometric, electronic and magnetic properties of fluorine adatom-doped graphene nanoribbons (GNRs) are fulfilled by various concentrations and adsorption positions of adatoms for both armchair and zigzag edge-terminated shapes. The two-typical widths that can wholly represent for all AGNR and ZGNR are included in the studying model, as shown in Figs. 6(a) and 6(b). The unit cells are described by the number of dimers lines and zigzag lines (N\(_A \)
   and N\(_Z \)) along \(y\). The lattice constants along periodic direction \(x\), respectively, are 
   \(3b\)
     and \( 2\sqrt 3 b \) (\(b\): C-C bond length) for AGNR and ZGNR. Besides, unnecessary dangling bonds in the GNRs are fully eliminated by hydrogen passivation. The geometric stability could be evaluated by the binding energy $E_b$ = ($E_{tot}$ - $E_{GNRs}$ - n$E_{A}$)/n, in which $E_{tot}$, $E_{GNRs}$ and $E_{A}$ account for the total energies of the combined system, pristine GNRs and the energy of all isolated adatoms; n displays for the number of adatoms, respectively. The fluorine adatoms-adsorbed GNRs can be effectively synthesized from experiment since the evaluated binding energy is lower than -2.5 eV (Table 2). The top site doping  results in the optimized structures of halogen adatoms-adsorbed GNRs regardless of any doping cases. The inhomogeneous buckled honeycomb lattice structures are revealed in the fluorine adatoms-doped GNRs among other halogen adatoms-adsorbed ones. C atoms closest to the F adatoms are deviated from the graphene plane, and  they are sensitively depended on concentrations and adsorption positions. For the single F adatom adsorption, the deviation of C atoms are \(0.053\) \AA \  and \(0.133\) \AA \  at the center and edge of GNR, respectively. They obviously increase with the adatom concentrations, e.g., \(0.204\) \AA \ for highest adsorption concentration. The shortest and longest F-C bond lengths (\(1.387\) \AA \ and \(1.55\) \AA \ ) correspond to the highest adsorption concentration and the ribbon center-positioned single F adsorption. Likewise, the C-C bond lengths nearest to F adatoms are extended in the range of ~0.05-0.11 \AA \ as compared with those of pristine GNRs. This demonstrates the obvious change of the critical chemical bonds owing to the significant fluorination effects. 
     
The honeycomb lattice symmetry, finite-size quantum confinement and edge structure lead to many special 1D energy bands in the pristine AGNR (Fig. 7(a)). Specifically, there exist asymmetry between the occupied valence bands and the unoccupied conduction bands around the Fermi level (\(E_F\)), where the direct band gap of \( E_g^d  = 0.6  \)  eV emerges from the effect of the finite-size confinement at the \(\Gamma \) point. The  \(\pi\) bonds of parallel 2p\(_z  \) orbitals, and the \(\sigma\) bonds of (2p\(_x\), 2p\(_y\)) orbitals are responsible for the electronic states of \( E^{c,v}  \le 2  \) eV and the deeper ones, as indicated in DOS (Fig. 11(a)).  Most of dispersions in the 1D band structures are parabolic, while few of them have partially flat ones. All the energy dispersions depend on wave vectors monotonously except for the subband anti-crossings.  The band-edge states which appear at \(k_x=0,1 \) (in unit of \( \pi /3b \)), and other states are related to subband anti-crossings that create the van Hove singularities in DOSs.

The  energy bands are dramatically changed due to the adatom doping effects. For the single F adatom-adsorbed AGNR, it presents the metallic behavior with or without spin splitting, as shown in Fig. 7(b) and 7(c). The Fermi level is shifted to valence energy bands, namely a red shift. The free hole densities exist in the unoccupied valence states between two Fermi momenta (\(\pm k_F \)  related to two valence bands intersecting with \(E_F\)). The \(\pi\) bonds of C atoms create the low-lying energy bands, independent of adatom adsorption positions. Furthermore, the (F, C) co-dominated energy bands, accompanied with the \(\sigma\) bands, occur at \( E^v  \le  - 2.5\) eV. They possess the weak energy dispersions or the narrow energy band widths. The critical F-C bonds and the significant modifications of \(\pi\) and \(\sigma\) bands can be characterized by the special features of 1D energy bands.

The variations in the concentration, adsorption position, and edge structure can create the obvious change of the 1D energy bands.  With the gradual rise of adatom concentration, the Fermi-momentum states are drastically varied, as shown in Figs. 7(d)-7(f) for two-F adsorptions. The total free hole densities become higher [\((5,\, \textcolor{red}{18})_d\) in Fig. 7(e); \((5,\, 18)_s\) in Table 2] or remain the same [\((2,\, \textcolor{red}{24})_d\) in Fig. 7(d); \((2,\, 24)_s\) in Table 2], compared to that of the single adatom adsorption (Figs. 7(b) and 7(c)). Note that expressions, [\((5,\, 18)_s\) and \((5,\, \textcolor{red}{18})_d\) or \((2,\, 24)_s\) and \((2,\, \textcolor{red}{24})_d\)], stand for the adatom adsorption positions under the single-side (black number; subscript s) and double-side (black and red number; subscript d) adsorption. Furthermore, the metallic 1D band structures might be thoroughly turned into the semiconducting one for the very close adatom-adatom distances, e.g., an indirect gap of \(E_g^i  = 0.64  \)  eV for the \((2,\, \textcolor{red}{5})_d\) adsorption (Fig. 7(f)). This exhibits the termination of the extended \(\pi\) bonds in AGNRs. Specifically, most of the F-doped AGNRs become the large-gap semiconductors under the efficiently high F adsorption concentrations (Figs. 7(g) and 7(h)). There exist more F-dominated energy bands which are located in the valence bands. The energy bands determine the magnitude of energy gap, and \( E_g \)  also depends on the \(\sigma\) bonding of carbon atoms (DOSs in Figs. 11(b) and 11(c)). Under the \(100\% \) adsorption, the nonmagnetic wide-gap semiconductor of \(E_g^i  = 3.2 \)  eV is generated (Fig. 7(h)).

A clear examination on the free hole densities is worthy of the detailed calculations. Such free hole densities are expressed by the relation \(\lambda  = 2k_F /\pi\), and linearly proportional to the Fermi momentum for each partially unoccupied valence band. There are no linear relations between \(\lambda\) and F adatom concentration except for the single F adatom adsorption. The single F adatom-doped AGNR can only create one free hole for any adsorption positions (Table 2), i.e., one electron is absorbed by a fluorine adatom from the bonded carbon atom due to the strong electron affinity. The metallic 4-, 6-, and 8-adatom adsorptions can create two or one free holes in a unit cell corresponding to the spin-degenerate and spin-split energy bands, respectively (Table 2). The F adatoms-adsorbed GNRs are in sharp contrast with the alkali adatoms-doped ones. The latter belongs to n-type metals even for the \(100\% \)   adatom concentration. Independent of adatom distributions, each alkali adatom generates one conduction electron from the outermost s orbital by means of the significant alkali-C bond.  This makes conduction electrons very high. 

The electronic and magnetic properties are diversified by the various edge structures. There exist a major difference in energy bands of AGNRs and ZGNRs. Pristine ZGNRs possess a pair of partially flat valence and conduction bands nearest to  \(E_F\) at small \(k_x\)'s (Fig. 8(a)), corresponding to wave functions localized at the zigzag boundaries.\cite{46} Moreover, these energy bands also exhibit the double degeneracy for the spin degree of freedom even if they are closely related to the anti-ferromagnetic configuration across the ribbon center and the ferromagnetic one at the same edge (Fig. 9(d)). Besides, the band-edge states of N\(_Z=8\) ZGNR create a direct gap of \(E_g^d  = 0.46\) eV. The partially flat 1D band structures with the localized charge distributions could be changed by F adatom doping such as energy dispersions, energy gap, and state degeneracy (Figs. 8(b)-8(f)). For the single F adatom-adsorbed ZGNRs, the number of edge-localized energy bands are reduced to a half of the spin splitting (Fig. 8(b)). Furthermore, the energy bands intersect with the Fermi level that exhibits the metallic behavior. Such behavior vanishes under the adsorption of  two-F adatoms which locate at both zigzag edges [\((3,\, 30)_s\) in Fig. 8(c) and \((3,\, \textcolor{red}{30})_d\) in Fig. 8(d)], leading to a direct-gap semiconductor (Table 2). Under the central adsorption of  two-F adatoms, the behavior of an indirect narrow-gap semiconductor is presented with spin splitting (Fig. 8(e)). Especially, the partially flat semiconducting band structures fully turn into the spin splitting metallic ones under the adsorption of two-F adatoms which situates at edge and center [\((3,\, \textcolor{red}{14})_d\) in Fig. 8(f)].  

 There exist five kinds of spin-dependent electronic and magnetic configurations. The pristine AGNRs are confinement-induced semiconductors without spin-split energy bands and magnetism (the first kind in Fig. 7(a) and Table 2). The similar semiconducting properties are also revealed in the F adatoms-doped AGNRs (Figs. 7(f)-7(h)). Nevertheless, the metallic systems might present the spin-degenerate energy bands without magnetism (the second kind in Figs. 7(b) and 7(e)), or the spin splitting with the ferromagnetic configuration (the third kind in Figs. 7(c) and 7(d)). On the other hand, the pristine ZGNRs are the anti-ferromagnetic semiconductors without spin splitting (the fourth kind in Fig. 8(a)). Furthermore, the F adatoms-doped ones present the first kind (Figs. 8(c) and 8(d)), the third kind (Figs. 8(b) and 8(f)), or the fifth kind (the semiconducting behavior with the anti-ferromagnetic configuration (Fig. 8(e))).

The electronic and magnetic configurations could be clearly interpreted by the spatial spin densities and magnetic moments.  Third, fourth and fifth kinds of the electronic and magnetic configurations exhibit the different arrangements (Figs. 9(a)-9(d), Fig. 9(e) and Fig. 9(f)), in which the competition between the spin-up and spin-down configurations create the net magnetic moment in a unit cell (Table 2). The spin densities are mostly distributed near the edge structures except for the fifth kind (Fig. 9(f)). When only one F adatom is adsorbed at the edge of AGNR, this leads to the ferromagnetic spin-up configuration, as shown in Fig. 9(a), and results in a net moment of 0.47 \(\mu _B  \). For two-F adsorptions near both boundaries, AGNRs exhibit the enhanced ferromagnetism across the ribbon center with 0.76 \(\mu _B  \) (Figs. 9(b) and 9(c)), while magnetism is absent in ZGNRs (Figs 9(e) and 9(f)). These illustrate that F adatoms close to the armchair and zigzag edges, respectively, create and destroy the same-spin arrangement on the spot. Specifically, two F adatoms adsorption at edge and center of ZGNR creates to an unusual anti-ferromagnetic configuration with a zero magnetic moment(Fig. 9(f)).

 The multi-orbital hybridizations are analyzed by the spatial charge density \((\rho )\), the charge density difference \((\Delta \rho )\), and the partial charge density  \(
                           (\rho_P )
                           \). \((\Delta \rho )\)  is calculated by subtracting the charge density of GNRs and F adatoms from that of F adatom-doped system. Also, the chemical bonds as well as the charge transfer are clearly illustrated by \((\rho ) \). The \( \pi \) and \( \sigma  \) bonds in the planar GNRs are formed by the parallel  2p\(_z\) orbitals and the planar (2p\(_x\), 2p\(_y\)) ones, as shown in Fig. 10(a) for \(N_A=12\) AGNR (the solid and dashed rectangles). The high charge density between F and C could have obviously changes in \( \pi \) bonding and observable reductions in \( \sigma  \) bonding (Figs. 10(b) and 10(d)). The strong fluorination effects are also revealed in drastic density variations, especial for \((\Delta \rho ), \)  near F adatoms on \((x,z)\) and \((y,z)\) planes (Figs. 10(c) and 10(e)). These clearly indicate the complicated (2p\(_x\), 2p\(_y\), 2p\(_z\))-(2p\(_x\), 2p\(_y\), 2p\(_z\)) hybridizations in F-C bonds. When the distance between two fluorine adatoms is sufficiently short, there exist the significant (2p\(_x\), 2p\(_y\)) hybridizations in F-F bonds, as illustrated in Figs. 10(c) and 10(e) (red rectangles on  \((x,z)\) and  \((y,z)\) planes). As for the metallic and semiconducting behaviors, they are characterized by the distorted \( \pi \) bondings in the partially charge density related to electronic states very close to \(E_F\) (Figs. 10(f) \(\&\) 10(h), and Figs. 10(g) \(\&\) 10(i)).

 The diverse electronic and magnetic configurations can be verified by the orbital-projected DOSs (Fig. 11). There exhibit many special structures, in which the asymmetric and symmetric peaks emerge from the parabolic and partially flat energy bands. When the adatom concentration is lower than \(
                                          50\%\) (Figs. 11(b), 11(d) and 11(e)), the low-energy DOS is dominated by the \( \pi \) bonding of C-2p\(_z\) orbitals (red curves). Such bonding also makes contributions to the deeper-energy DOS. The peak structures owing to the \( \sigma  \) bonding of (2p\(_x\), 2p\(_y\)) orbitals occur at \(
                                           E <  - 2.5\)  eV (green curves). The similar features take place with the respect to the (2p\(_x\), 2p\(_y\)) and 2p\(_z\) orbitals of the F adatoms (dashed blue and pink curves). The former possess a sufficiently wide energy width of 2 eV, so that there exist the significant (2p\(_x\), 2p\(_y\)) orbital hybridizations in F-F bonds. All the orbitals can create the merged peak structures at deeper energy, clearly illustrating the ((2p\(_x\), 2p\(_y\), 2p\(_z\))-(2p\(_x\), 2p\(_y\), 2p\(_z\))) multi-orbital hybridizations in F-C bonds. Specifically, the energy width of the bands might be more than 5 eV under the highest adatom concentration (Figs. 11(c)). Furthermore, the corresponding peaks are stronger than those of the \( \sigma  \) bands. The occupied valence bands are closely related to the (2p\(_x\), 2p\(_y\)) orbitals of F and C, especially for the orbital hybridization in F-F bonds.
                                           
Five kinds of electronic and magnetic configurations are characterized by the specific peak structures near the Fermi level. A pair of anti-symmetric peaks near \(E=0\), which are divergent in the opposite direction, represent an energy gap in the absence of spin splitting (the first kind), as shown in Figs. 11(a)-11(c). But for non-magnetic metals (the second kind), the similar pair presents a blue shift (\(0.5 - 1.0\))  eV (Figs. 11(d)), and DOS is finite at the Fermi level. The spin-polarized peak structures are revealed in ferromagnetic metals (the third kind in Figs. 11(e) and 11(g)). The low-energy peaks are quite different for spin-up and spin-down configurations, in which they are asymmetric about \(E=0\), and the former predominates the occupied states of \(E<0\). Specifically, the partially flat energy bands in a pristine ZGNR can create a pair of symmetric peaks (blue circles in Fig. 11(f)), accompanied with an energy gap and one peak due to an extra band-edge state (Fig. 8(a)). This corresponds to an anti-ferromagnetic semiconductor with spin degeneracy (the fourth kind). There are more pairs of symmetric peaks centered about \(E=0\)  in the presence of spin splitting, as indicated in Fig. 11(h). The fifth kind of peak structure arises from the narrow-gap semiconducting band structure of two F adatoms-absorbed ZGNRs with the anti-ferromagnetic behavior.

\subsection{3. Experimental measurements}

On the experimental aspects, a scanning tunneling microscope (STM), an useful tool for imaging surfaces at atomic level with a good resolution, has been successfully utilized to identify the unique geometric structures of the graphene-related systems, including graphite, graphene, graphene compounds, carbon nanotubes, and GNRs.  The atomic-scaled observations clearly reveal the 2D networks of local defects,\cite{47} the buckled and rippled structures of graphene islands,\cite{48} the adatom distributions on graphene surface,\cite{49} the nanoscale width of GNR,\cite{50} and the chiral arrangements of the hexagons on the planar edges\cite{51} and a cylindrical surface.\cite{52} For fluorinated 2D and 1D graphene, the buckled structure, the adatom height, and the bond length (Table 1 and 2) are worthy of further STM verifications. Such measurements are very useful in identifying the significant multi-orbital hybridizations in F-C, F-F and C-C chemical bonds.

Angle-resolved photoemission spectroscopy (ARPES), a direct experimental technique to observe the distribution of electrons in the reciprocal space of solids, is one of the most powerful experimental methods to examine the wave-vector-dependent electronic structures. Besides, the feature-rich electronic structures within the distinct dimensions are revealed in the experimental measurements on graphene-related systems. For example, the verified energy bands include an isotropic Dirac-cone structure with linear energy dispersions in monolayer graphene,\cite{53} two pairs of parabolic bands in bilayer AB stacking,\cite{54} the bilayer- and monolayer-like energy dispersions, respectively, at \(k_z=0\) and zone boundary in AB-stacked graphite\cite{55} and 1D parabolic energy bands with a direct energy gap in AGNRs.\cite{56} In addition, an edge-localized partially flat band is deduced to have an association with the zigzag-like steps on graphite surface. The measurement on fluorinated graphene shows that the Fermi level is shifted 0.79 eV below the Dirac point, indicating the p-type doping. Up to now, the ARPES measurements on the adatom-enriched 1D energy bands of GNRs are absent. The ARPES and spin-resolved ARPES\cite{57} are available in verifying five kinds of predicted electronic and magnetic configurations in F adatom-absorbed GNRs.

The scanning tunneling spectroscopy (STS), an extension of STM, is used to provide information about the density of electrons in a sample as a function of their energy, with the tunneling differential conductance \((dI/dV)\) proportional to DOS, could serve as very efficient methods to identify the dimension-enriched special structures in DOS.  The measured DOSs show the splitting \(\pi\) and 
                                 \(
                                 \pi ^ *\)   peaks and a finite value near  \(E_F\) characteristic of the semi-metallic behavior in graphite,\cite{58} a linear E-dependence vanishing at the Dirac point in monolayer graphene, the asymmetry-created peak structures in bilayer graphene,\cite{59} a prominent peak at \(E_F\) arising from partially flat bands in tri-layer and penta-layer ABC-stacked graphene.\cite{60} The geometry-dependent energy gaps and the asymmetric peaks of 1D parabolic bands in carbon nanotubes\cite{61} and GNRs. The STS and spin-resolved STS could be utilized to examine the orbital-projected DOSs of F adatom-absorbed 2D graphene and 1D GNRs, covering the finite value at \(E_F\), energy gap and spin-polarized peak structures.

\section{IV. Concluding remarks}

 The geometric structures, electronic and magnetic properties of F adatom-adsorbed 2D graphene and 1D GNRs are investigated using the first-principles calculations. Discussions on the  adatom-dominated band structure, the spin arrangement, spatial charge density, and the orbital-projected density of states (DOS) are useful to elucidate the feature-rich electronic and magnetic properties. The similar analyses could be further generalized to the emergent 2D and 1D layered materials, with nanoscale thickness and unique lattice symmetries, covering 2D and 1D silicene, germanene, tinene, phosphorene, MoS\(_2\) and so on. The fluorination-induced diverse effects mainly arise from the complicated relations among 2D and 1D honeycomb lattice symmetries, the top site F-C multi-orbital hybridizations, finite-size quantum confinements and specific edge structures. The metallic (with/without spin-splitting) or large-gap semiconducting behaviors can turn into the highly potential applications, such as electronic, optical, and spintronic devices.
           
           Each F adatom-adsorbed 2D and 1D graphene can create an inhomogeneous buckled structure, in which the adatom height and the change of C-C bond length are mainly determined by the significant fluorination effect. The multi-orbital hybridizations in F-C, C-C and F-F bonds of fluorinated graphene lead either to the metals (with/without magnetism) or to the large-gap semiconductors (without magnetism), depending on the adatom concentrations and distributions. The carbon-dominated magnetisms are revealed in the fluorination of 12.5\% and 16.7\%. Furthermore,  the critical multi-orbital hybridizations, and the edge-dependent spin distributions of the fluorinated GNRs are responsible for five kinds of electronic and magnetic properties, covering the non-magnetic and ferromagnetic metals, non-magnetic semiconductors, and anti-ferromagnetic semiconductors with/without the spin splitting. Besides, F adatom-doped AGNRs could create one or two holes per unit cell corresponding to the spin-split and spin-degenerate \(\pi\)-electronic energy bands (non-magnetism and ferromagnetism), respectively. However, F adatom-doped ZGNRs only exhibit the case with lower carrier density. The predicted geometric, electronic and magnetic properties of fluorinated 2D and 1D graphene could be verified by STM, ARPES and STS, respectively.

  \par\noindent {\bf Acknowledgments}
            
            This research is funded by Vietnam National Foundation for Science and Technology Development (NAFOSTED) under grant number 103.01-2017.74. Also, we would like to thank Department of Physics, National Cheng Kung University, Taiwan for the computational support.
\bibliography{achemso}

\newpage \centerline {\Large \textbf {FIGURE CAPTIONS}}

\vskip0.5 truecm 
\begin{itemize}
\item[Figure 1:] Geometric structures with top and side views for various concentrations and distributions: (a) F:C = 8:8 = 100\% (double-side), (b) F:C = 2:8 = 25\% (para-site), (c) F:C = 2:8 = 25\% (meta-site), (d) F:C = 1:6 = 16.7\%, (e) F:C = 1:8 = 12.5\%, and (f) F:C = 2:18 = 11.2\% (para-site).
\bigskip

\item[Figure 2:] Band structures of various concentrations: (a) pristine graphene, (b) F:C = 8:8 = 100\% (double-side), (c) F:C = 2:8 = 25\% (para-site), (d) F:C = 1:6 = 16.7\%, (e) F:C = 1:8 = 12.5\%, (f) F:C = 2:18 = 11.2\% (para-site). The green and purple circles correspond to the contributions of F and passivated C atoms, respectively.
\bigskip

\item[Figure 3:] The spin-density distributions with top and side views for: (a) F:C = 16.7\%, and (b) F:C = 12.5\%, . The red isosurfaces represent the charge density of spin-up configuration.
\bigskip

\item[Figure 4:] The spatial charge densities for: (a) pristine graphene, (b) F:C = 1:8 = 12.5\%, and (c) F:C = 1:6 = 16.7\%. The corresponding charge density differences are, respectively, shown in (d) \& (e).
\bigskip

\item[Figure 5:] Orbital-projected DOS for: (a) pristine graphene, (b) F:C = 8:8 = 100\% (double-side), (c) F:C = 2:8 = 25\% (para-site), (d) F:C = 1:6 = 16.7\%, (e) F:C = 1:8 = 12.5\%, (f) F:C = 2:18 = 11.2\% (para-site).

\item[Figure 6:] Geometric structures of F-adsorbed GNRs for (a) N\(_A=12\) armchair and (b) N\(_Z=8\) zigzag systems. The red rectangles represent the unit cells. The lattice constants are, respectively, \(a=3b\)  and 
                \(
                a = 2\sqrt 3 b
                \)
                  for armchair and zigzag GNRs. Numbers on the top of carbons denote the positions of adatoms.

\item[Figure 7:] Band structures of N\(_A=12\) AGNR for (a) pristine, (b) \((13)\)-, (c) \((2)\)-, (d) \((2,\textcolor{red}{24})_d\)-, (e) \((5,\textcolor{red}{18})_d\)-, (f) \((2,\textcolor{red}{5})_d\)-, (g) (10F)\(_d\)-, (h)
                                     (24F)-adsorption; Violet circles represent the contribution of F adatoms. The red and black curves denote the spin-split energy bands. [\((2, 24)_s\) and \((2,\textcolor{red}{24})_d\)] stand for the  single-side (black number; subscript s)  and double-side (black and red number; subscript d) adsorption positions,  (10F)\(_d\) and (24F)\(_d\) represent the double-side adsorption with 10 and 24 adatoms, respectively.

\item[Figure 8:] Band structures of N\(_Z=8\) ZGNR for (a) pristine, (b) \((14)\)-, (c) \((3, 30)_s\)-, (d) \((3,\textcolor{red}{30})_d\)-, (e) \((10,\textcolor{red}{14})_d\)-, \(\&\) (f) \((3,\textcolor{red}{14})_d\)-adatom adsorption; Violet circles represent the contribution of F adatoms. The red and black curves denote the spin-split energy bands. The subscripts $s$ and $d$, respectively, correspond to the single- and double-side adsorptions, accompanied with the black number and (black, red) numbers.

\item[Figure 9:] Spin density of N\(_A=12\) AGNR for (a) \((2)\)-, (b) \((2, 24)_s\)-, (c) \((2,\textcolor{red}{24})_d\)-adatom adsorption, and N\(_Z=8\) ZGNR for (d) \((14)\)-, (e) pristine, and (f) \((10,\textcolor{red}{14})_d\)-adatom adsorption.

\item[Figure 10:] Spatial charge density of N\(_A=12\) AGNR for (a) Pristine, (b) \((13)\)-, \(\&\) (d) (24F)\(_d\)-adatom adsorption; Charge density difference of N\(_A=12\) AGNR for (c) \((13)\)-, (e) (24F)\(_d\)-adatom adsorption. Partial charge density is shown for (f) AGNR-\((5,\textcolor{red}{18})_d\)-, (g) ZGNR-\((3,\textcolor{red}{14})_d\)-, (h) AGNR-\((2,\textcolor{red}{5})_d\)-, and (i) ZGNR-\((10,\textcolor{red}{14})_d\)-adatom adsorption. 

\item[Figure 11: ] Orbital-projected DOSs for (a) Pristine, (b) (10F)\(_d\)-, (c) (24F)\(_d\)-, (d) \((13)\)-, (e) \((2,\textcolor{red}{24})_d\)-adsorbed N\(_A=12\) AGNR;  (f) pristine;  (g) \((14)\)-, (h) \((10,\textcolor{red}{14})_d\)-adsorbed N\(_Z=8\) ZGNR. Blue circles correspond to the partially flat bands. 
\end{itemize}

\newpage
\begin{table}
\caption{The calculated C-C and C-F bond lengths, heights (shift on z-axis) of passivated carbons, binding energies, energy gaps, and total magnetic moments per unit cell of fluorinated graphene systems. The double-side structures are labeled with *}
\label{table}
\centering
\begin{tabular*}{\textwidth}{@{\extracolsep{\fill}}llllllllll}
\hline
Adatom& F:C&\% & &Bond length& &Height &$E_b$ &$E_g$&M$_{tot}$\\
 & & &C-F (\AA)&1st C-C (\AA)&2nd C-C (\AA)& (\AA)&(eV)& (eV)&($\mu_B$)\\
\hline
F& 8:8$^*$ & 100 &1.38 &1.58 & &0.48& -3.00 &3.12  & 0 \\	
 & 4:8$^*$ (para) & 50 &1.48 &1.51 &  & 0.38  & -2.71  &  2.77 &0\\
 & 4:8 (meta) & 50 &1.48 &1.5 &  & 0.28 & -1.16  & 0  &1.94\\
 & 2:8$^*$ (para) & 25 & 1.48 & 1.52 &1.41  & 0.36 & -2.96 & 2.72 &0\\	
 & 2:8$^*$ (meta) & 25 &1.54&  1.48&1.43  &0.34 & -2.68  & 0  &0\\	
 & 1:6 & 16.7 & 1.53 & 1.48& 1.42&  0.32& -2.45&  0&0.22\\	
 & 1:8 & 12.5 & 1.53 & 1.49 & 1.42 &0.32  &-2.64  & 0 &0.5\\
 & 2:18$^*$ (para) & 11.2 &1.48  &1.51  &1.42  &  0.34&-2.46 &  0&0\\	
\hline
\end{tabular*}
\end{table}

\begin{table}[htb]
                    \caption{Binding energy, Bohr magneton and magnetism, energy gap, number of holes in a unit cell, and geometric parameters for N\(_A=12\) armchair and N\(_Z=8\) zigzag GNRs under single- and double-side adsorption. NM, FM and AFM correspond to non-magnetism, ferro-magnetism and anti-ferro-magnetism, respectively.}
                         \label{table2}
                           \begin{center}
                                                
                           \begin{tabular}{ |l|l|l|l|l|l|l|l|l|}
                                \hline
                                                GNRs & \makecell{Adsorption\\ configurations }  & \(E_b\) (eV)
                                                 & \makecell{Bohr\\ magneton\\(\(\mu _B \) )/\\magnetism } & \makecell{ \( E_g^{d(i)} \) \\(eV)/ \\Metal}  & \makecell{Num-\\ber\\of\\holes} & \makecell{F-C\\ (\AA)}  & \makecell{C \\deviation\\(\AA)}  & \makecell{Nearest\\C-C\\(\AA)}  \\ 
                                                \hline
                                                \multirow{21}{4em}{AGNR\\N\(_A=12\)} & Pristine & \(0\) & 0/NM &  \( E_g^{d}=0.6 \)& 0 & 0 & 0 & 1.43 \\ 
                                                 & \((13)\) & \(-3.0145\) & 0/NM & M & 1 & 1.55 & 0.05 & 1.48 \\ 
                                                 & \((2)\) & \(-2.5155\) & 0.47/FM & M & 1 & 1.47 & 0.13 & 1.49 \\ 
                                                 & \((5, 18)_s\) & \(-2.6246\) & 0/NM & M & 2 & 1.55 & 0.055 & 1.49 \\ 
                                                 & \((5,\textcolor{red}{18})_d\) & \(-3.0145\) & 0/NM & M & 2 & 1.55 & 0.054 & 1.49 \\ 
                                                 & \((2, 24)_s\) & \(-3.3314\) & 0.76/FM & M & 1 & 1.47 & 0.132 & 1.49 \\ 
                                                 & \((2,\textcolor{red}{24})_d\) & \(-3.3526\) & 0.76/FM & M & 1 & 1.47 & 0.131 & 1.49 \\ 
                                                 & \((2,\textcolor{red}{5})_d\) & \(-4.2646\) & 0/NM & \( E_g^{i}=0.64 \) & 0 & 1.443 & 0.161 & 1.493 \\ 
                                                & (C:\textcolor{red}{F}=24:\textcolor{red}{4})\(_d\) & \(-3.1538\) & 0/NM & M & 2 & 1.503 & 0.097 & 1.491 \\ 
                                               &   (C:\textcolor{red}{F}=24:\textcolor{red}{4})\(_d\) & \(-3.3412\) & 0.56/FM & 	M & 1 & 1.454 & 0.15 & 1.497 \\ 
                                              
                                              & (C:\textcolor{red}{F}=24:\textcolor{red}{4})\(_d\) & \(-4.2922\) & 0/NM & \( E_g^{i}=0.65 \) & 0 & 1.43 & 0.17 & 1.496 \\ 
                                                 & (C:\textcolor{red}{F}=24:\textcolor{red}{6})\(_d\) & \(-3.0107\) & 0/NM & M & 2 & 1.532 & 0.068 & 1.497 \\ 
                                                 & (C:\textcolor{red}{F}=24:\textcolor{red}{6})\(_d\)& \(-3.1625\) & 0.57/FM & M & 1 & 1.442 & 0.162 & 1.49 \\ 
                                                 & (C:\textcolor{red}{F}=24:\textcolor{red}{6})\(_d\) & \(-3.9762\) & 0/NM & \( E_g^{i}=0.66 \) & 0 & 1.413 & 0.185 & 1.51 \\ 
                                                 & (C:\textcolor{red}{F}=24:\textcolor{red}{8})\(_d\)  & \(-3.4174\) & 0/NM & M & 2 & 1.446 & 0.154 & 1.5 \\ 
                                                 & (C:\textcolor{red}{F}=24:\textcolor{red}{8})\(_d\)  & \(-3.3418\) & 0.59/FM & M & 1 & 1.472 & 0.127 & 1.5 \\ 
                                                 & (C:\textcolor{red}{F}=24:\textcolor{red}{8})\(_d\) & \(-3.8643\) & 0/NM & \( E_g^{i}=0.72 \) & 0 & 1.424 & 0.182 & 1.51 \\ 
                                                 & (C:\textcolor{red}{F}=24:\textcolor{red}{10})\(_d\) & \(-3.8426\) & 0/NM & \( E_g^{i}=0.85 \) & 0 & 1.425 & 0.182 & 1.51 \\ 
                                                 & (C:\textcolor{red}{F}=24:\textcolor{red}{14})\(_d\) & \(-3.8521\) & 0/NM & \( E_g^{i}=2.25 \) & 0 & 1.412 & 0.188 & 1.53 \\ 
                                                 & (C:\textcolor{red}{F}=24:\textcolor{red}{20})\(_d\) & \(-3.5124\) & 0/NM & \( E_g^{i}=2.69 \) & 0 & 1.417 & 0.193 & 1.53 \\ 
                                                 & (C:\textcolor{red}{F}=24:\textcolor{red}{24})\(_d\) & \(-3.6153\) & 0/NM & \( E_g^{d}=3.2 \) & 0 & 1.387 & 0.204 & 1.54 \\ 
                                                \hline
                                                \multirow{6}{4em}{ZGNR\\N\(_Z=8\)} & Pristine & \(0\) & 0/AFM & \( E_g^{d}=0.46 \) & 0 & 0 & 0 & 1.43 \\ 
                                                 & \((3)\) & \(-4.5388\) & 0.42/FM & M & 1 & 1.46 & 0.139 & 1.49 \\ 
                                                 & \((14)\)  & \(-2.5693\) & 0.4/FM & M & 1 & 1.55 & 0.06 & 1.48 \\ 
                                                 & \((3, 14)_s\) & \(-3.2856\) & 0.4/FM & M & 1 & 1.45 & 0.15 & 1.49 \\ 
                                                 & \((3,\textcolor{red}{14})_d\) & \(-3.4725\) & 0.4/FM & M & 1 & 1.45 & 0.15 & 1.49 \\ 
                                                 & \((10,\textcolor{red}{14})_d\) & \(-3.5521\) & 0/AFM & \( E_g^{i}=0.2 \) & 0 & 1.51 & 0.09 & 1.49 \\ 
                                                 & \((3, 30)_s\) & \(-4.2989\) & 0/NM & \( E_g^{i}=0.46 \) & 0 & 1.55 & 0.146 & 1.49 \\ 
                                              
                                                 & \((3,\textcolor{red}{30})_d\) & \(-4.6528\) & 0/NM & \( E_g^{d}=0.46 \) & 0 & 1.46 & 0.145 & 1.49 \\ 
                                                  & (C:\textcolor{red}{F}=32:\textcolor{red}{28})\(_d\) & \(-3.3645\) & 0/NM & \( E_g^{d}=2.78 \) & 0 & 1.38 & 0.219 & 1.52 \\ 
                                                 & (C:\textcolor{red}{F}=32:\textcolor{red}{32})\(_d\) & \(-3.4547\) & 0/NM & \( E_g^{d}=3.1 \) & 0 & 1.38 & 0.219 & 1.52 \\ 
                                                \hline
                                                \end{tabular}
                                                 \end{center}
                                                 \end{table}

\newpage
\begin{figure}[!h]
\centering
\includegraphics[height=20cm]{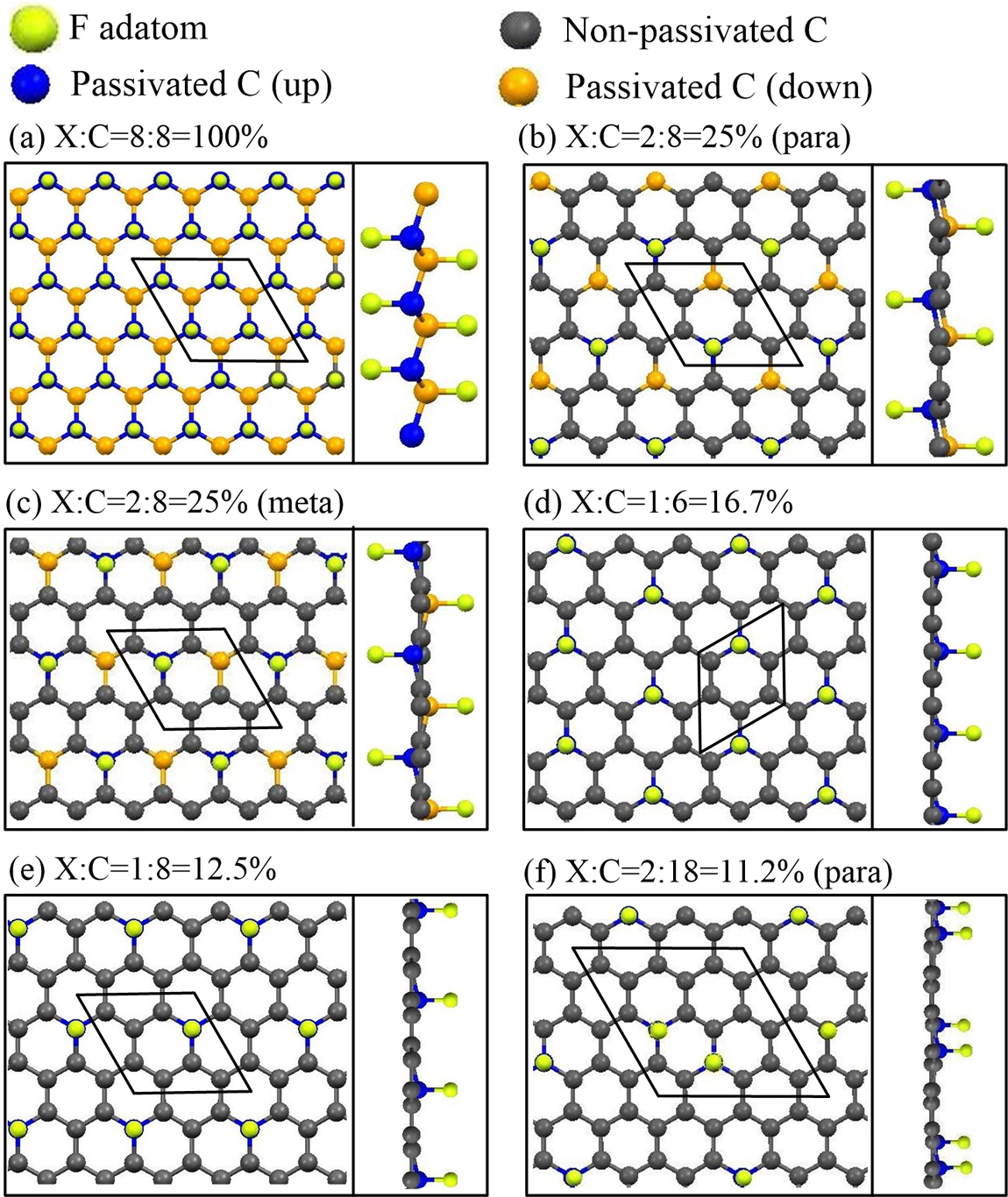}
\caption{Geometric structures with top and side views for various concentrations and distributions: (a) F:C = 8:8 = 100\% (double-side), (b) F:C = 2:8 = 25\% (para-site), (c) F:C = 2:8 = 25\% (meta-site), (d) F:C = 1:6 = 16.7\%, (e) F:C = 1:8 = 12.5\%, and (f) F:C = 2:18 = 11.2\% (para-site).}
\end{figure}

\newpage
\begin{figure}[!h]
\centering
\includegraphics[height=20cm]{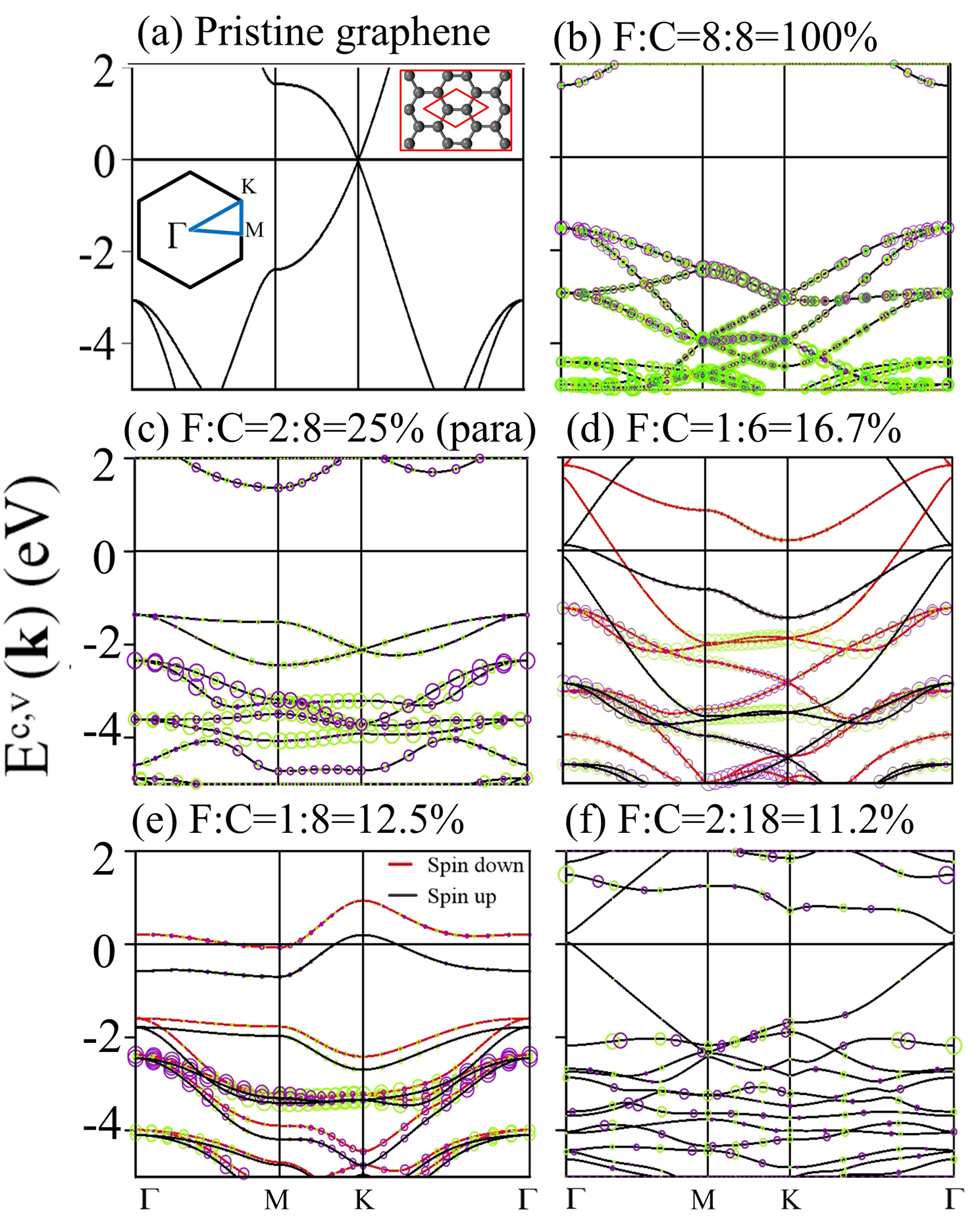}
\caption{Band structures of various concentrations: (a) pristine graphene, (b) F:C = 8:8 = 100\% (double-side), (c) F:C = 2:8 = 25\% (para-site), (d) F:C = 1:6 = 16.7\%, (e) F:C = 1:8 = 12.5\%, (f) F:C = 2:18 = 11.2\% (para-site). The green and purple circles correspond to the contributions of F and passivated C atoms, respectively.}
\end{figure}

\newpage
\begin{figure}[!h]
\centering
\includegraphics[height=8cm]{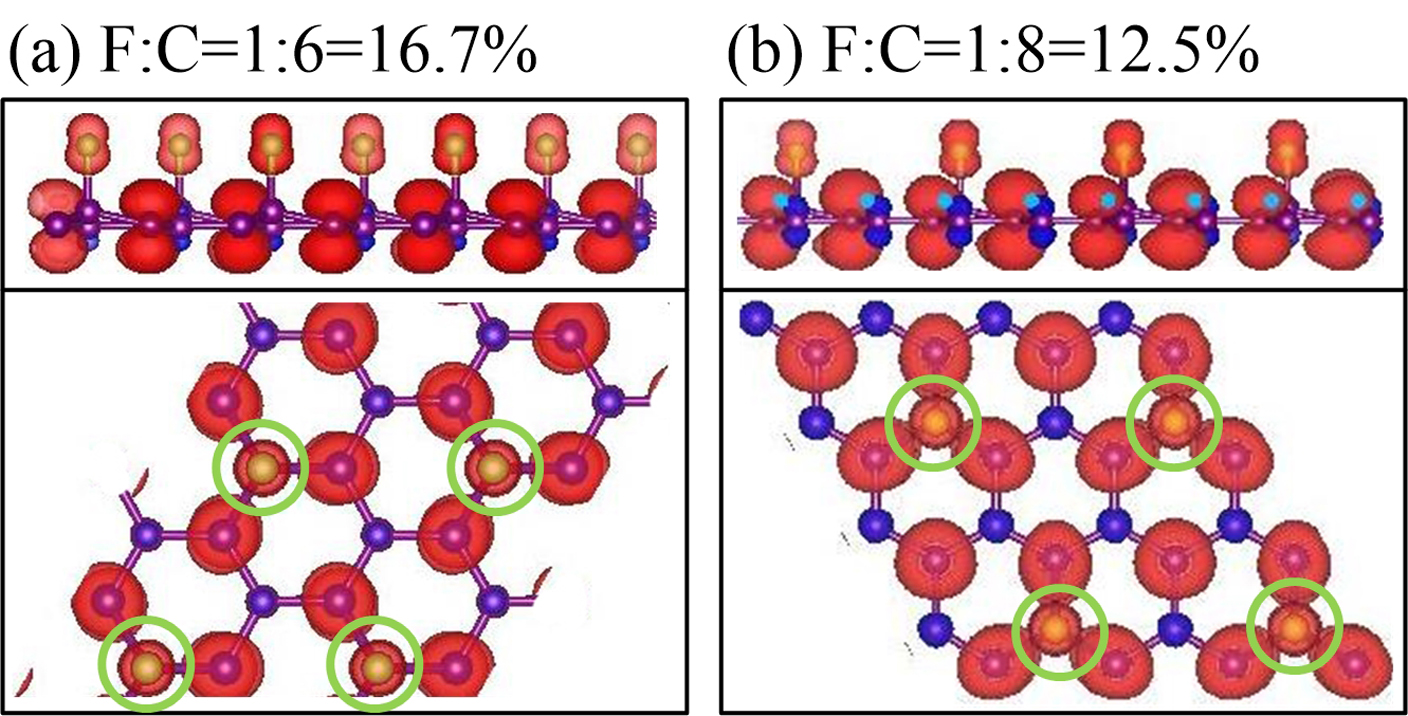}
\caption{The spin-density distributions with top and side views for: (a) F:C = 16.7\%, and (b) F:C = 12.5\%, . The red isosurfaces represent the charge density of spin-up configuration.}
\end{figure}

\newpage
\begin{figure}[!h]
\centering
\includegraphics[width=11cm, height=20cm]{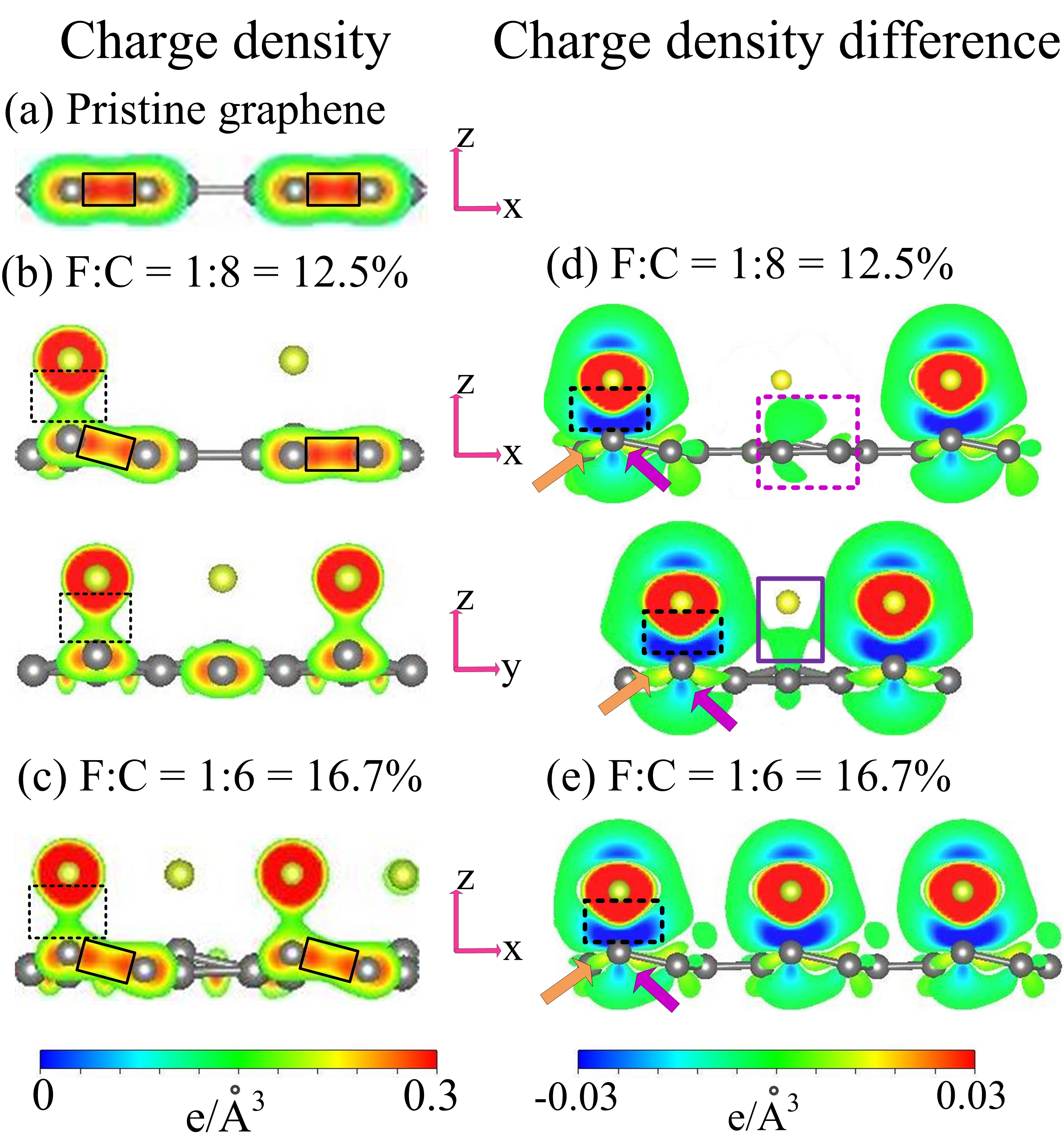}
\caption{The spatial charge densities for: (a) pristine graphene, (b) F:C = 1:8 = 12.5\%, and (c) F:C = 1:6 = 16.7\%. The corresponding charge density differences are, respectively, shown in (d) \& (e).}
\end{figure}

\newpage
\begin{figure}[!h]
\centering
\includegraphics[height=20cm]{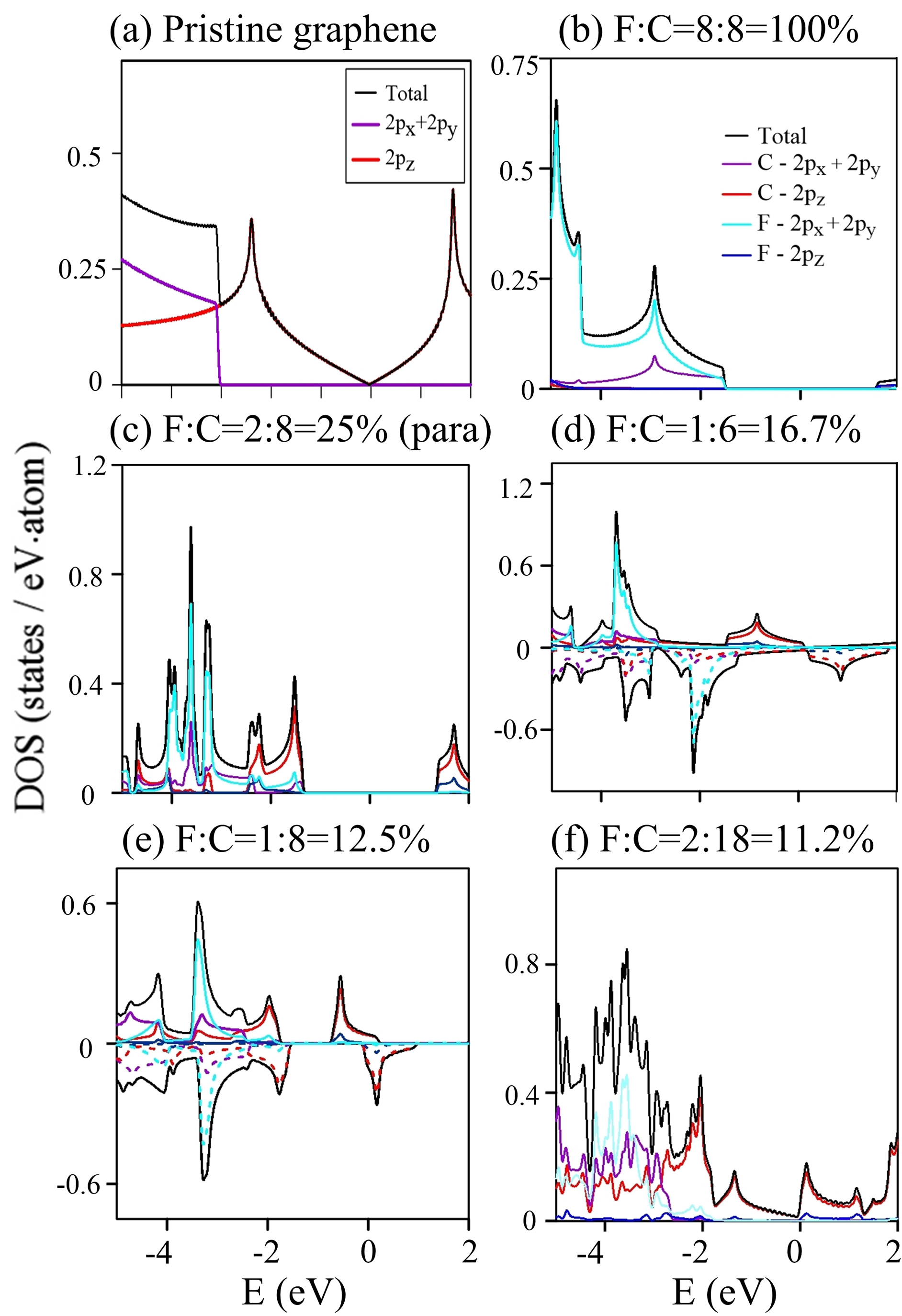}
\caption{Orbital-projected DOS for: (a) pristine graphene, (b) F:C = 8:8 = 100\% (double-side), (c) F:C = 2:8 = 25\% (para-site), (d) F:C = 1:6 = 16.7\%, (e) F:C = 1:8 = 12.5\%, (f) F:C = 2:18 = 11.2\% (para-site).}
\end{figure} 
\newpage
\begin{figure}[htb]
                \includegraphics[width=8cm, height=20cm]{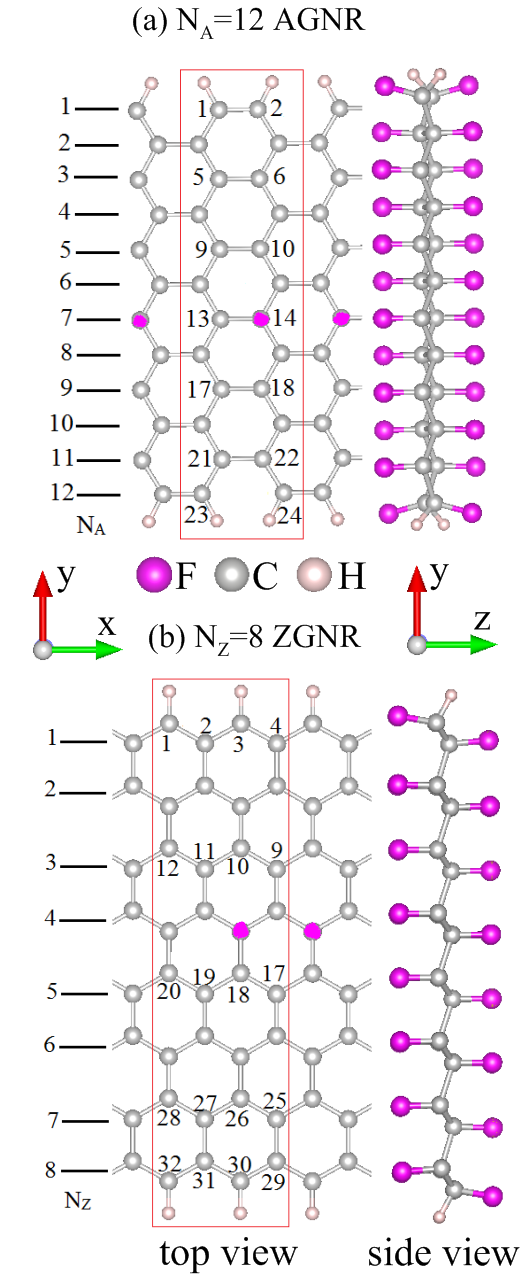}
                \caption{Geometric structures of F-adsorbed GNRs for (a) N\(_A=12\) armchair and (b) N\(_Z=8\) zigzag systems. The red rectangles represent the unit cells. The lattice constants are, respectively, \(a=3b\)  and 
                \(
                a = 2\sqrt 3 b
                \)
                  for armchair and zigzag GNRs. Numbers on the top of carbons denote the positions of adatoms.}
                \label{fgr:6}
              \end{figure}

\newpage
\begin{figure}[htb]
                          \includegraphics[width=7.5cm, height=20cm]{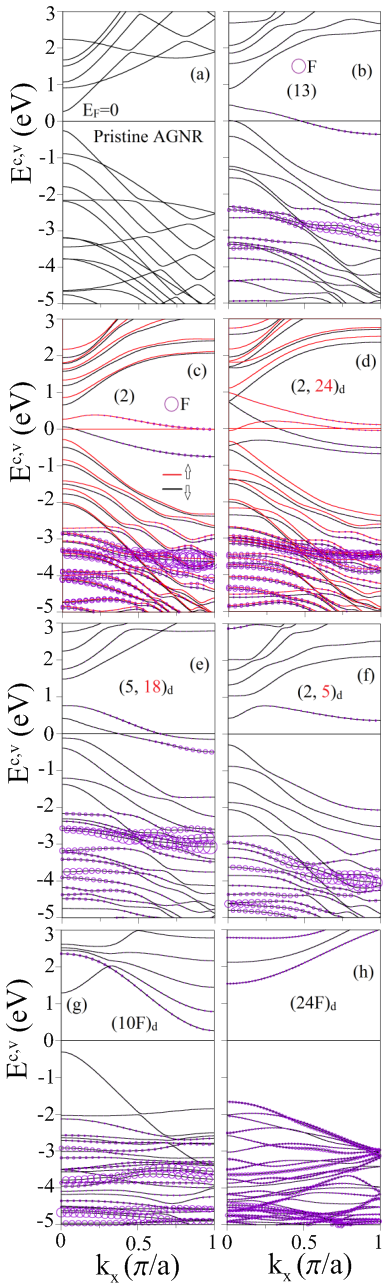}
                                     \caption{Band structures of N\(_A=12\) AGNR for (a) pristine, (b) \((13)\)-, (c) \((2)\)-, (d) \((2,\textcolor{red}{24})_d\)-, (e) \((5,\textcolor{red}{18})_d\)-, (f) \((2,\textcolor{red}{5})_d\)-, (g) (10F)\(_d\)-, (h)
                                     (24F)-adsorption; Violet circles represent the contribution of F adatoms. The red and black curves denote the spin-split energy bands. [\((2, 24)_s\) and \((2,\textcolor{red}{24})_d\)] stand for the  single-side (black number; subscript s)  and double-side (black and red number; subscript d) adsorption positions,  (10F)\(_d\) and (24F)\(_d\) represent the double-side adsorption with 10 and 24 adatoms, respectively.}
                                     \label{fgr:7}
                               \end{figure} 
 \newpage
 \begin{figure}[htb]
                              \includegraphics[width=8cm, height=18cm]{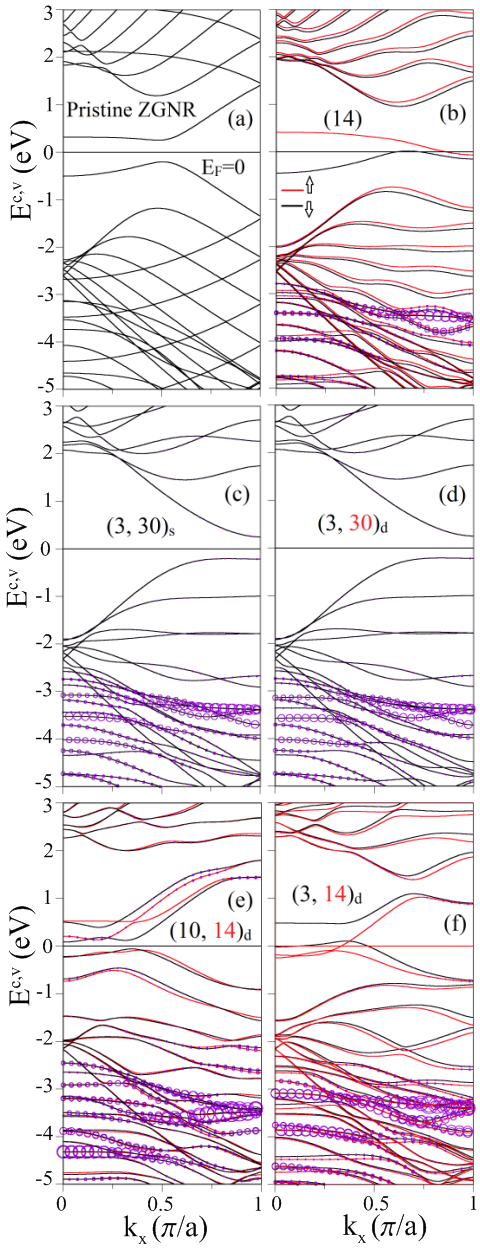}
                               \caption{Band structures of N\(_Z=8\) ZGNR for (a) pristine, (b) \((14)\)-, (c) \((3, 30)_s\)-, (d) \((3,\textcolor{red}{30})_d\)-, (e) \((10,\textcolor{red}{14})_d\)-, \(\&\) (f) \((3,\textcolor{red}{14})_d\)-adatom adsorption; Violet circles represent the contribution of F adatoms. The red and black curves denote the spin-split energy bands. The subscripts $s$ and $d$, respectively, correspond to the single- and double-side adsorptions, accompanied with the black number and (black, red) numbers.}
                              \label{fgr:8}
                              \end{figure}    
                
\newpage
\begin{figure}[htb]
                              \includegraphics[width=10cm, height=17cm]{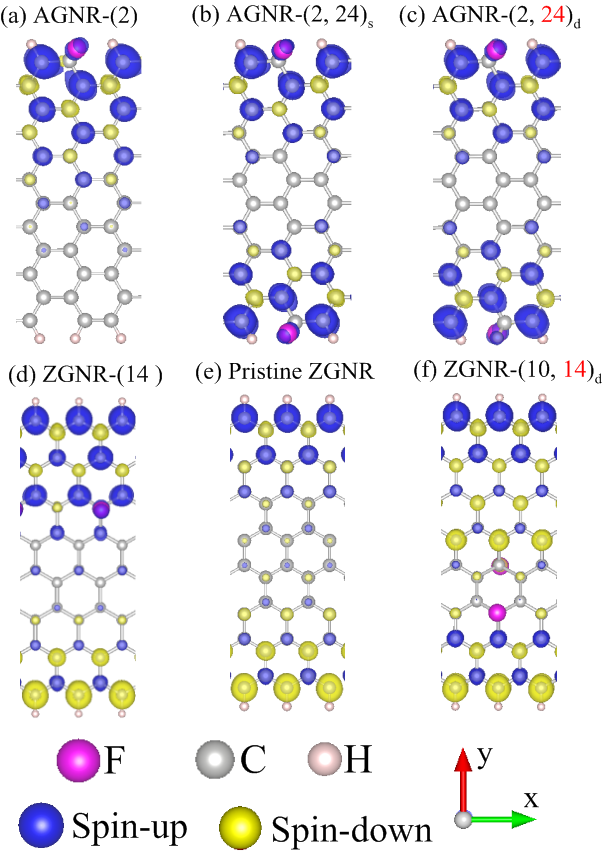}
                               \caption{Spin density of N\(_A=12\) AGNR for (a) \((2)\)-, (b) \((2, 24)_s\)-, (c) \((2,\textcolor{red}{24})_d\)-adatom adsorption, and N\(_Z=8\) ZGNR for (d) \((14)\)-, (e) pristine, and (f) \((10,\textcolor{red}{14})_d\)-adatom adsorption.
                               }
                               \label{fgr:9}
                               \end{figure}
\newpage
\begin{figure}[htb]
                               \includegraphics[width=10cm, height=18cm]{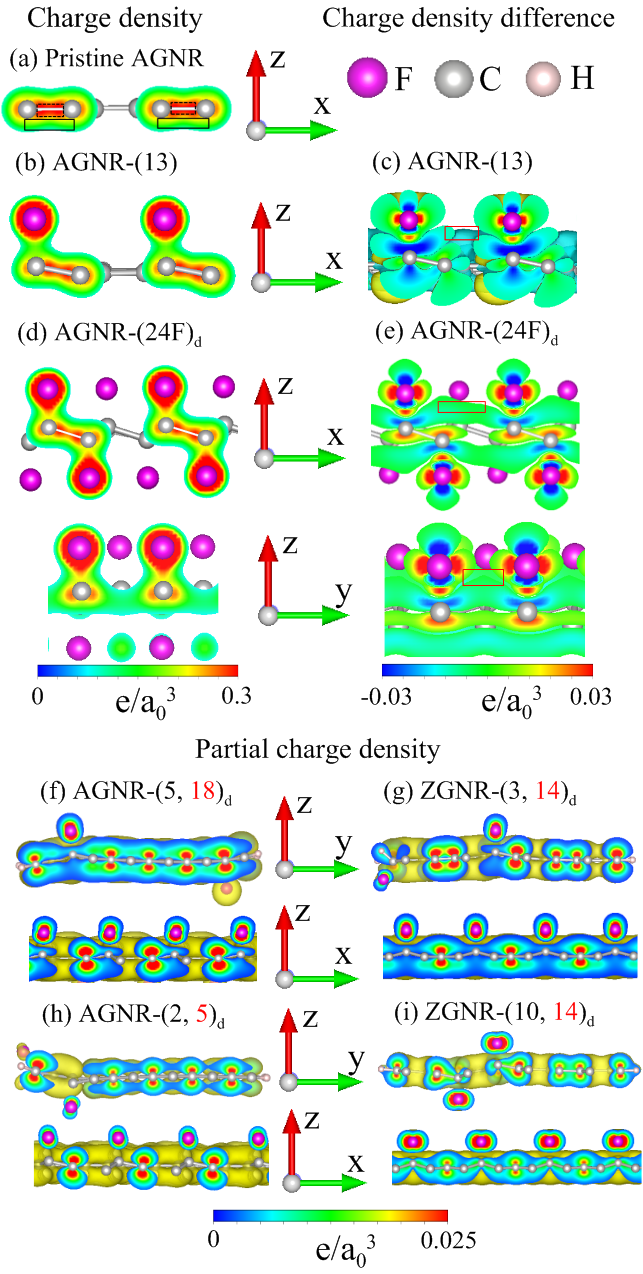}
                                 \caption{Spatial charge density of N\(_A=12\) AGNR for (a) Pristine, (b) \((13)\)-, \(\&\) (d) (24F)\(_d\)-adatom adsorption; Charge density difference of N\(_A=12\) AGNR for (c) \((13)\)-, (e) (24F)\(_d\)-adatom adsorption. Partial charge density is shown for (f) AGNR-\((5,\textcolor{red}{18})_d\)-, (g) ZGNR-\((3,\textcolor{red}{14})_d\)-, (h) AGNR-\((2,\textcolor{red}{5})_d\)-, and (i) ZGNR-\((10,\textcolor{red}{14})_d\)-adatom adsorption.} 
                                     \label{fgr:10}
                              \end{figure}        

\newpage
\begin{figure}[htb]
                                 \includegraphics[width=10cm, height=18cm]{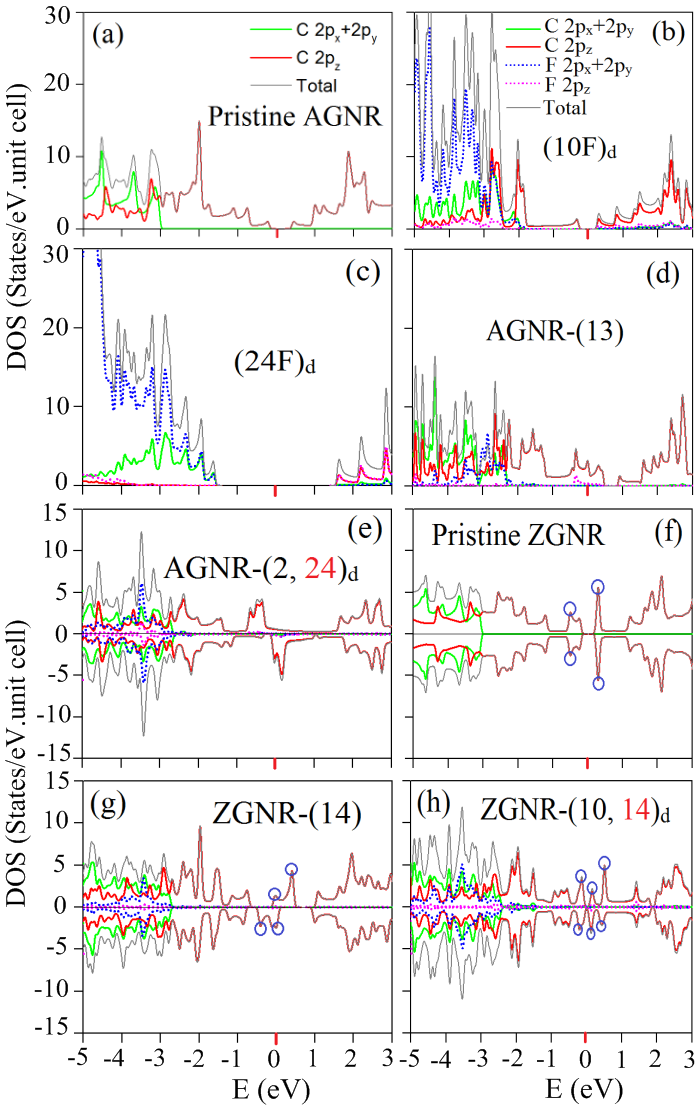}
                                  \caption{Orbital-projected DOSs for (a) Pristine, (b) (10F)\(_d\)-, (c) (24F)\(_d\)-, (d) \((13)\)-, (e) \((2,\textcolor{red}{24})_d\)-adsorbed N\(_A=12\) AGNR;  (f) pristine;  (g) \((14)\)-, (h) \((10,\textcolor{red}{14})_d\)-adsorbed N\(_Z=8\) ZGNR. Blue circles correspond to the partially flat bands.}
                                   \label{fgr:11}
                                  \end{figure}    

\end{document}